\documentclass[10pt,aps,prd,twocolumn,nofootinbib,showpacs,showkeys,superscriptaddress]{revtex4-2}

\usepackage{psfrag}
\usepackage{mathrsfs}
\usepackage{amssymb, bm}
\usepackage{amsmath, amsthm}
\usepackage{epstopdf}
\usepackage[breaklinks=true]{hyperref}
\usepackage{enumerate}
\usepackage{longtable}
\usepackage{subfigure}
\usepackage{color}
\usepackage{mathrsfs}
\usepackage{graphicx}
\usepackage{bm,natbib,url,textcase}
\usepackage{xurl}
\usepackage[title]{appendix}

\usepackage[english]{babel}
\usepackage[T1]{fontenc}
\usepackage{comment}
\numberwithin{equation}{section}

\usepackage{geometry}
\allowdisplaybreaks

\hypersetup{
 colorlinks=true, 
 linkcolor=blue, 
 citecolor=blue, 
 urlcolor=blue 
}

\usepackage{tikz}

\newcommand{\be}{\begin{equation}}
\newcommand{\ee}{\end{equation}}

\definecolor{purple}{rgb}{1,0,1}
\definecolor{lime}{HTML}{a6CE39} 

\newcommand{\orcidicon}{%
	\begin{tikzpicture}
		\draw[lime, fill=lime] (0,0) 
		circle [radius=0.15] 
		node[white] {{\fontfamily{qag}\selectfont \tiny ID}};
		\draw[white, fill=white] (-0.0625,0.095) 
		circle [radius=0.007];
	\end{tikzpicture}	\hspace{-2mm}
}

\newcommand\orcidMarcello{{\href{https://orcid.org/0000-0003-0397-2705}{\orcidicon}}}
\newcommand\orcidMax{{\href{https://orcid.org/0000-0003-0325-3911}{\orcidicon}}}
\newcommand\orcidValerio{{\href{https://orcid.org/0000-0002-2601-1870}{\orcidicon}}}

\DeclareMathOperator{\arctanh}{arctanh}

\def\z{~} 
\def\nn{\nonumber} 
\def\T{{\cal T}}
\def\K{{\cal K}}

\begin{document}
\def\theequation{\arabic{section}.\arabic{equation}} 

\title{How the Schwarzschild--de Sitter horizons remain in thermal 
equilibrium at vastly different temperatures}

\author{Marcello Miranda\orcidMarcello}
\email{marcello.miranda@unina.it}
\affiliation{Scuola Superiore Meridionale, Largo San Marcellino 10, 
I-80138, Napoli, Italy}
\affiliation{INFN Sez. di Napoli, Compl. Univ. Monte S. Angelo, 
Edificio G, Via	Cinthia, I-80126, Napoli, Italy}

\author{Massimiliano Rinaldi\orcidMax}
\email[]{massimiliano.rinaldi@unitn.it}
\affiliation{Department of Physics, University of Trento, Via 
Sommarive 14, 38122 Trento, Italy} 
\affiliation{Trento Institute for Fundamental 
Physics and Applications TIFPA-INFN, Via 
Sommarive 14, 38122 Trento, Italy} 

\author{Valerio Faraoni\orcidValerio} 
\email[]{vfaraoni@ubishops.ca}
\affiliation{Department of Physics \& Astronomy, Bishop's University, 2600 
College Street, Sherbrooke, Qu\'ebec, Canada J1M~1Z7}

\begin{abstract}

The Tolman\textendash{}Ehrenfest criterion of thermal equilibrium for a static fluid in a static spacetime is generalized to stationary heat conduction, in the approximation in which backreaction is negligible. 
Applying this generalized criterion to the Hawking radiation in the Schwarzschild--de Sitter geometry shows that the two horizons (which act as thermostats) remain in thermal equilibrium. The temperature of the radiation fluid interpolates between the temperatures at the horizons, with a static analytic profile that is given explicitly.

\end{abstract}


\maketitle
\section{Introduction}
\label{sec:1}
\setcounter{equation}{0}

It is well established that black holes are not black, but radiate Hawking radiation when quantum fields are present \cite{Hawking:1975vcx}. 
A Schwarzschild black hole with mass $m$ and line element\footnote{We follow the notation of Ref.~\cite{Wald:1984rg}.} 
\be
ds^2 =-\left( 1-\frac{2m}{r} \right) dt^2 +\frac{dr^2}{ 1-2m/r} + r^2 
d\Omega_{(2)}^2 
\ee
(where $d\Omega_{(2)}^2 \equiv d\vartheta^2 +\sin^2 \vartheta \, d\varphi^2$ is the metric on the unit 2-sphere) has a horizon at $r=2m$ radiating at the Hawking temperature 
$\T_\mathrm{H}=1/\left( 8\pi m 
\right)$ \cite{Hawking:1975vcx}. Similarly, the de Sitter geometry 
\be
ds^2 =-\left(1-\frac{\Lambda r^2}{3} \right) dt^2 
+\frac{dr^2}{1-\Lambda r^2/3} + r^2 
d\Omega_{(2)}^2 
\ee
(where $\Lambda>0$ and $H=\sqrt{\Lambda/3}$ are, respectively, the 
cosmological and the Hubble constants) has a null horizon at $r=1/H$, 
radiating at the Gibbons\textendash{}Hawking temperature $\T_\mathrm{GH} = H/ (2\pi )$ \cite{Gibbons:1977mu}. A geometry interpolating between the two above is the so-called  Schwarzschild--de Sitter/Kottler (SdS) spacetime, with metric
\begin{align}
ds^2&=-f(r)dt^2+\frac{dr^2}{f(r)}+r^2d\Omega_{(2)}^2\,,\label{SdS}\\
 f(r)&=1-\frac{2m}{r}-\frac{\Lambda r^2}{3} \,.
\end{align}
The metric is a solution of the vacuum Einstein equation with cosmological constant $\Lambda$ and can have two horizons, according to the values of $m$ and $\Lambda$. Here, we assume that the black hole radius is much smaller than the cosmological horizon size, i.e., $ 2m \ll H^{-1} =\sqrt{3/\Lambda} $. In this approximation, $\T_\mathrm{H} \approx {1}/(8\pi m) \gg \T_\mathrm{GH} \approx {H}/(2\pi)$. Then the temperatures of the two horizons are vastly different, with the black hole horizon being much hotter than the de Sitter one. It is often stated in the literature that, because of the temperature difference between the black hole and cosmological horizons, there cannot be thermal equilibrium in Schwarzschild-de Sitter (SdS) space—except for the degenerate Nariai case \cite{Nariai}, where the two horizons coincide (a scenario which is not physically interesting).
Consequently, multiple temperatures must be introduced, and some authors argue that the area law or other fundamental relations of black hole thermodynamics must be modified (e.g., \cite{Choudhury:2004ph, Aros:2008ef, Saida:2009ss, Kim:2014zta, Pappas:2017kam, Robson:2019yzx, Nakarachinda:2021jxd, Volovik:2023stf}). These conflicting proposals point to an incomplete understanding of multi-horizon spacetimes in general, and of SdS in particular. Here we propose an alternative view in which the two horizons remain in thermal equilibrium although their temperatures are different. Thermal equilibrium (defined as the time-independence of temperature and heat flux in the static frame) is achieved by letting the temperature vary with the radial position, interpolating between the Hawking and the Gibbons\textendash{}Hawking temperatures (we provide explicitly the analytical temperature profile $\T(r)$ between the two horizons in the approximation of negligible back-reaction). Here by ``temperature'' we mean the temperature of the blackbody Hawking radiation between the two horizons, which will be treated as a radiation fluid at rest in the locally static geometry between these horizons. Since backreaction is neglected in the calculation of both the Hawking and the Gibbons\textendash{}Hawking temperatures \cite{Hawking:1975vcx, Gibbons:1977mu}, the test fluid approximation for the blackbody radiation is legitimate in the ``usual'' regime of moderate emission. The temperature profile is derived from a generalization of the Tolman\textendash{}Ehrenfest criterion \cite{Tolman28, Tolman:1930zza, Tolman:1930ona} for the thermal equilibrium of a test fluid at rest in a static spacetime. This generalized criterion applies in the presence of heat sources, which are absent in the ``standard'' situation conceived by Tolman. 

To begin with, we recall that, long ago, Tolman \cite{Tolman28, Tolman:1930zza}, and then Tolman and Ehrenfest \cite{Tolman:1930ona}, showed that a test fluid at rest in a static spacetime in general relativity cannot be in thermal equilibrium at constant temperature. There must necessarily be a temperature gradient (unless $g_{00}$ is constant), and the Tolman\textendash{}Ehrenfest criterion for thermal equilibrium is \cite{Tolman28, Tolman:1930zza,Tolman:1930ona}
\be
\T \sqrt{-g_{00}}= \mbox{const.} \label{standardTolman}
\ee
in coordinates adapted to the time symmetry. The physical interpretation of this result is that heat is a form of energy; therefore, it sinks in a gravitational field and spacetime regions where gravity is stronger are hotter than those where gravity is weaker. The modern derivation of the Tolman\textendash{}Ehrenfest criterion \cite{MTW, Santiago:2018kds, Santiago:2018jeu, Santiago:2019aem} uses Eckart's law of heat conduction (one of the three constitutive relations assumed in Eckart's first-order thermodynamics \cite{Eckart40_I, Eckart:1940te, Klein49}) for dissipative fluids. For the reader's convenience, we report this derivation in Appendix~\ref{Appendix:A}.

A relativistic dissipative fluid has a stress-energy tensor of the form \cite{Eckart:1940te}
\begin{align}\label{imperfect}
 T^{ab} & = \rho u^{a}u^{b}+Ph^{ab}+q^{a}u^{b} + q^{b} u^{a} 
+\pi^{ab}\,,
\end{align}
where $u^{a}$ is the timelike fluid 4-velocity (normalized to $u_{a}u^{a}=-1$), 
\be
h_{ab} \equiv g_{ab}+u_{a} u_{b} 
\ee
is the metric induced onto the 3-space orthogonal to $u^c$, 
\be
\rho = T^{ab} u_{a}u_{b} 
\ee 
is the energy density, 
\be
P= \frac{1}{3} \, T^{ab}h_{ab}
\ee 
is the isotropic pressure, 
\be 
q^{a}=-T^{cd} \, u_{c}h_{d}{}^{a}
\ee 
is the heat flux density, 
\be
\pi^{ab}= h^{a}{}_{c} \, h^{b}{}_{d} \, T^{cd}-P h^{ab}
\ee 
is the (traceless) anisotropic stress tensor. $h_{ab}$, $q_a$, and $\pi_{ab}$ are purely spatial according to the observer $u^c$, 
\be
 h_{ab} u^a = h_{ab} u^b = q_a u^a = \pi_{ab}u^a = \pi_{ab}u^b = 0 \,.
\ee
According to Eckart's law (generalizing the Fourier law familiar from pre-relativistic physics), the heat flux density is \cite{Eckart:1940te} 
\be
q^a =- \K\, h^{a b } \left( \nabla_b \T + \T \dot{u}_b \right) \,,
\ee
where $\K$ is the thermal conductivity. In the standard treatment, thermal equilibrium is defined by the absence of heat flow $ q^a = 0 $.

The conditions for the validity of the Tolman\textendash{}Ehrenfest criterion are quite restrictive, and it is not surprising that the latter has seen very limited applications. There are, however, applications to material physics \cite{Luttinger:1964zz} and to neutron star cores \cite{Laskos-Patkos:2022lgy,Li:2022url,Kim:2021kou,Kim:2022qlc}, see \cite{Santiago:2018kds,Santiago:2018jeu,Santiago:2019aem} for a revisitation from first principles and \cite{Lima:2019brf,Lima:2021ccv} for further discussion. More recently, the Tolman\textendash{}Ehrenfest criterion has been extended to conformally static spacetimes \cite{Faraoni:2023gqg} and to scalar-tensor gravity \cite{Karolinski:2024ukr}.

Equation~(\ref{standardTolman}) makes it clear that a temperature gradient is necessary for thermal equilibrium and therefore thermal equilibrium between the two horizons of the SdS spacetime is not ruled out {\em a priori}. In fact, we will show that it is possible. To this end, we require thermal equilibrium for the radiation fluid generated by the two static horizons, and we apply a generalized Tolman\textendash{}Ehrenfest criterion to it. The criterion~(\ref{standardTolman}) must be extended because it was derived in the absence of sources, while in the SdS case the two horizons act as thermostats keeping these two boundaries at fixed temperatures. To understand the physics at play, it is useful to make an analogy with heat conduction in a metal rod of finite length (indeed, Eckart's law is just a generalization of the non-relativistic Fourier law of heat conduction $\vec{q}=-\K \vec{\nabla} \T$).

In the absence of heat sources, if the ends of the rod are initially at different temperatures heat flows from the hotter to the cooler regions until the temperature gradient $\vec{\nabla} \T$ disappears, thermal equilibrium is established, the heat flux density $\vec{q}=-\K \vec{\nabla} \T$ vanishes, and the two ends reach the same temperature, constant across the rod and time-independent. (In the relativistic case, the gravitational field instead maintains a temperature gradient, as discussed above.) But one can also consider {\em stationary heat conduction}, in which the two ends of the rod are kept at different temperatures $\T_1$ and $\T_2$ (constant in time) by thermostats. Then, heat flows from the hotter to the cooler end establishing a linear temperature profile, with $\T(x)$ and $\vec{q}$ independent of time. The one-dimensional heat equation
\be
\frac{\partial \T}{\partial t} =a \, \frac{\partial^2 \T}{\partial x^2} 
\ee
(where $a$ is the Fourier coefficient), degenerates into $d^2 \T/dx^2 =0$. This situation is more similar to 
the case of the radiation fluid between the two horizons of SdS, which act as thermostats. To make the analogy more appropriate, instead of a one-dimensional rod one should consider stationary heat conduction in a three-dimensional medium with spherical symmetry, purely radial heat flow, and spherical thermostats. 

The next section discusses the ``standard'' Tolman\textendash{}Ehrenfest criterion, which is generalized to stationary heat conduction and applied to the SdS geometry in Sec.~\ref{sec:3}, where the temperature profile between the two horizons is also computed. Section~\ref{sec:4} discusses the regularization of this temperature at the horizons, while Sec.~\ref{sec:5} contains the conclusions.

\section{Standard Tolman\textendash{}Ehrenfest criterion and dissipative fluids}
\label{sec:2}
\setcounter{equation}{0}

In the derivation of the ``standard'' Tolman\textendash{}Ehrenfest condition for the thermal equilibrium of a static test fluid in a static spacetime, in the absence of heat sources or sinks \cite{MTW, Santiago:2018kds, Santiago:2018jeu, Santiago:2019aem}, one uses a dissipative fluid at rest. 
The velocity gradient is decomposed as 
\be
\nabla_{b}u_{a} =\sigma_{ab}+ \frac{\Theta}{3} \,h_{ab}+\omega_{ab}
-\dot{u}_{a}u_{b} \,, 
\ee
where an overdot denotes differentiation along the fluid 4-velocity, $\dot{u}_{a} \equiv u^{c}\nabla_{c} u_a $. The kinematic quantities of the fluid are the expansion scalar 
\be
 \Theta\equiv\nabla_{a}u^a \,,
\ee 
the trace-free shear tensor
\be
 \sigma^{ab} \equiv \left(h^{ac} h^{bd} 
-\frac{1}{3} \, h^{cd}h^{ab} \right) \nabla_{(c} u_{d)} \,,
\ee 
and the vorticity tensor 
\be
\omega^{ab} \equiv h^{ac} h^{bd} \nabla_{[d} u_{c]} \,.
\ee 
The velocity gradient is decomposed into symmetric and antisymmetric parts, and the former into trace-free and pure trace parts. The covariant derivative operator can be split into a contribution parallel to the 4-velocity and one perpendicular to it, 
\be
 \nabla_a ={h_a}^{b} \nabla_{b} -u_a \, u^b \nabla_b \,.
\ee

The dissipative nature of the fluid is described by the heat flux, the viscous pressure, and the anisotropic stresses. 

In Eckart's first-order thermodynamics \cite{Eckart:1940te}, an imperfect fluid at rest reduces to a perfect fluid because Eckart's theory for dissipative fluids is built assuming the three constitutive relations: 
\begin{eqnarray}
 q^a &=& -\K\,h^{a b } \left( \nabla_b \T + \T\dot{u}_b \right)\,, 
\label{Eckartheat}\\
&&\nonumber\\
P &=& P_{\rm perfect} + P_{\rm visc} = P_{\rm perfect} -\zeta\,\Theta 
\,,\\
&&\nonumber\\
\pi_{ab} &=& -2\,\eta\,\sigma_{ab} \,,
\end{eqnarray}
where $\zeta$ and $\eta$ are the bulk and shear viscosity coefficients, respectively. Apart from the inertial term $-\K\,\T\dot{u}_a$ in the heat flux density~(\ref{Eckartheat}) \cite{Eckart:1940te}, these constitutive relations are familiar from the 3-dimensional hydrodynamics of non-relativistic dissipative fluids with Newtonian fluid behavior. When the spatial velocity $u^i$ ($i=1,2,3$) vanishes, the dissipative fluid trivially reduces to a perfect one. In the absence of heat sources/sinks, thermal equilibrium is understood as the absence of heat flow $q^a=0$, from which one deduces the Tolman\textendash{}Ehrenfest condition of thermal equilibrium $\T \sqrt{-g_{00}} =$~const. beginning from Eckart's law~(\ref{Eckartheat}) (Appendix~\ref{Appendix:A}).

\section{Tolman\textendash{}Ehrenfest criterion generalized to stationary heat conduction}
\label{sec:3}
\setcounter{equation}{0}

Here we generalize the Tolman\textendash{}Ehrenfest criterion for thermal equilibrium to stationary heat conduction, in which $q^a\neq 0$ but $q^a$ and the fluid temperature $\T$ does not depend on time in the static frame.

Neglecting the backreaction of the quantum radiation onto the black hole and the de Sitter horizon, the equation of motion $\nabla^b T_{ab}=0$ of the dissipative fluid~(\ref{imperfect}) is projected\footnote{A complete description of energy conservation for the radiation fluid would take into account the fact that Hawking radiation from the black hole horizon pumps energy into the fluid (while the black hole mass decreases). This energy travels by conduction to the de Sitter horizon. The backreaction would change the blackbody spectrum and cause the temperature to evolve in time, effects that are beyond the scope of the present work. Therefore, covariant energy conservation $\nabla^b T_{ab}=0$ is only an approximation.} onto the time direction $u^a$, giving \cite{Ellis71}
\begin{equation}
 \dot{\rho}+\left( P+\rho \right) \Theta 
+ \pi_{ab}\sigma^{ab} + \nabla_a q^a + q_a \dot{u}^a =0 \,,\label{questa1}
\end{equation}
while the projection onto the 3-space orthogonal to $u^a$ yields
\begin{align}
 \left( P + \rho \right) \dot{u}_a + 
{h_a}^c \left(\dot{q}_c + \nabla_c P + \nabla^b \pi_{bc} \right) & \nn\\ 
+ \left( {\omega_a}^b + {\sigma_a}^b \right)q_b 
+\frac{4\Theta}{3}\, q_a & = 0 \,. \label{questa2}
\end{align}

For a static fluid at rest in a static spacetime, in coordinates adapted to the time symmetry, we have $\dot{\rho} \equiv u^c\nabla_c \rho = \partial \rho/\partial t = 0$, $\dot{P}=0$, $\pi_{ab}= \sigma_{ab}= \omega_{ab}=0$, and 
\be
\Theta = 
\frac{1}{\sqrt{-g}} \, \partial_a \left( \sqrt{-g} \, u^a\right) = 
\frac{1}{\sqrt{-g}} \, \partial_t \left( \sqrt{-g} \, u^0 \right) = 
\partial_t u^0 =0 \,,
\ee
where we used the fact that $u^0=\sqrt{ -g_{00}}$ (from the normalization $u^c u_c=-1$) and the time-independence of the metric components in adapted coordinates. As our new definition of thermal equilibrium for stationary heat conduction, we assume a steady heat flow, then ${h_a}^b \dot{q}_b=0 $ (see Appendix~\ref{Appendix:B}) and Eqs.~(\ref{questa1}) and (\ref{questa2}) reduce to 
\begin{align}
 \nabla_c q^c + q_a \dot{u}^a = 0 \,, \label{star}\\
		\nn \\
 \left( P+\rho \right) \dot{u}_a + {h_a}^c \nabla_c P = 0 \,.
\end{align}
Using the Buchdahl relation $\dot{u}_a= \nabla_a \left( \ln  \sqrt{-g_{00}} \, \right) $ valid in static spacetimes \cite{Buchdahl49} and $\nabla_a q^{a}= \partial_a \left( \sqrt{-g} \, q^a \right)/ \sqrt{-g}$, we have
\begin{align} 
\frac{1}{\sqrt{-g}} \, \partial_a \left( \sqrt{-g} \, q^a \right) = & -q^a \nabla_a \left( \ln \sqrt{-g_{00}} \, \right) \,,\label{Eq.1}\\
\nn &\\ 
\nabla_a \left( \ln{ \sqrt{-g_{00}}} \, \right) 
= & - \frac{\nabla_a P}{ P+\rho} \,.\label{Eq.2}
\end{align} 
Substituting Eq.~(\ref{Eq.2}) into~(\ref{star}) gives
\be
\frac{1}{ \sqrt{-g} } \, \partial_i \left( \sqrt{-g} \, q^i \right) 
- \frac{ q^j \nabla_j P }{ P+\rho}=0 \quad\quad \left( i,j=1,2,3 \right). 
\label{heat1}
\ee

For a static and spherically symmetric geometry with $q^a= \left( 0, q^r, 0, 0 \right)$, Eq.~(\ref{heat1}) reads
\be
\frac{1}{r^2\, q^r } \, \frac{d}{dr} \left( r^2 \, q^r 
\right) = \frac{1}{P+\rho} \, \frac{dP}{dr} = \frac{w}{(1+w)\rho} \, 
\frac{d\rho}{dr} \,, \label{heat3}
\ee
where we used the linear equation of state $P=w\rho$. Equation~(\ref{heat3}) has the general solution
\be
q^r(r) = \frac{C }{\sqrt{-g/\sin^2\vartheta}}\,\rho^{w/(1+w)} \,,\label{heat_rho}
\ee
with $C$ an integration constant. Notice that $\sin^2\vartheta$ removes the angular dependence of the metric determinant.

Equation~(\ref{Eq.2}) with the linear equation of state gives
\be
\nabla_a \left( \ln \sqrt{-g_{00}} \, \right) +\frac{w}{(w+1)\rho} \, 
\frac{d\rho}{dr}=0 \,,
\ee
which is immediately integrated to
\be
\rho(r)=\rho_0\,\left(\frac{1}{\sqrt{-g_{00}}}\right)^{\left(w+1\right)/{w}}\,.
\ee
with $\rho_0$ an integration constant, and, comparing with Eq.~(\ref{heat_rho}), one obtains
\be
q^r(r) = \frac{ q_0}{\sqrt{g_{00}\,g/\sin^2\vartheta}} \,.
\ee

where $q_0$ is an integration constant. Notice that $g/\sin^2\vartheta$ only depends on the radial coordinate. Now, let us specify the discussion to the SdS geometry~(\ref{SdS}) with $g=-r^2\sin^2{\vartheta}$, which produces 
\be
q^r (r) = \frac{q_0}{r^2 \sqrt{ -g_{00} }} =\frac{q_0}{r^2 \sqrt{ 1-2m/r - H^2r^2 }}\,. \label{3.13} 
\ee
{$q^r$ diverges on the horizons due to the adopted coordinate system. However, the Tolman criterion is necessarily formulated in this coordinate system, which is adapted to the time symmetry.}
The radial heat flux must be outgoing, which implies that $q_0>0$. 

Imposing now Eckart's constitutive relation~(\ref{Eckartheat}) gives 
\begin{align}
q^r &= \K\, g_{00} \left[ \frac{d\T}{dr} + \T \, \frac{d}{dr} \left( \ln 
\sqrt{-g_{00} } \, \right) \right]
\end{align}
which, combined with Eq.~(\ref{3.13}), yields
\be
\frac{d}{dr} \left( \T\sqrt{-g_{00}} \,\right) - \frac{q_0}{ \K\, r^2 g_{00}}=0 \,.
\ee
Assuming $\K=$~const., this equation integrates to 
\be
\T \sqrt{-g_{00}} = \T_0- \frac{q_0}{\K} \int \frac{dr}{ r^2 \left(1-2m/r 
-\Lambda r^2/3\right) } \,.
\ee 
being $\T_0$ a positive integration constant. As a check, this solution reduces to the ``standard'' Tolman\textendash{}Ehrenfest criterion~(\ref{standardTolman}) if the heat flux vanishes, $q_0 = 0$. 
The integral on the right-hand side is computed explicitly, giving the static local temperature profile 
\begin{widetext}
\be
\T (r) = \frac{1}{\sqrt{1-\frac{2m}{r} -\frac{\Lambda}{3}r^2}} 
\Bigg[ \T_0 + \frac{q_0}{2 m \K} \Bigg( \ln x 
 - \, \sum_{i} \, \frac{\left( \Lambda R_{i}^2 - 3 \right)\ln 
\left( x-x_i \right) }{ 3\left( \Lambda R_i^2 -1 \right)}\Bigg) 
\Bigg]\,, \label{Tprofile}
\ee
\end{widetext}
where $x\equiv r/r_0$, $r_0$ is a constant, $R_i$ are the roots of the equation $f(r)=0$, and $x_{i} = R_{i}/r_0$.

\subsection{Schwarzschild black hole}

For the pure Schwarzschild geometry with $\Lambda=0$ and $ f(r)=1-2m/r$, the temperature profile~(\ref{Tprofile}) reads
\begin{align}
 \T (r) = & \frac{1}{\sqrt{1-\frac{2m}{r}}} 
\Bigg[ \T_0-\frac{q_0}{2 m \K} \ln \left( 1-\frac{2m}{r} \right) \Bigg] 
\,. \label{mainresult}
\end{align}
Since the argument of the logarithm is smaller than unity, the requirement of a radially outgoing heat flow $q_0>0$ guarantees that the temperature $\T (r)$ is positive everywhere. As $r\to +\infty$, $\T $ must reduce to the Hawking temperature seen by the static distant observer, which identifies the integration constant $\T_0$ as
\be
\T_0= \frac{1}{8\pi m} \,.
\ee

The temperature $\T (r) $ diverges at the Schwarzschild horizon $r\to 2m$. This well-known fact is discussed in the next section. Similarly, the temperature~(\ref{Tprofile}) diverges at both the black hole and the de Sitter horizons.

\subsection{de Sitter space}
In the pure de Sitter case $m = 0$, the formal solution is 
\be
\T_\mathrm{dS} (r) = \frac{1}{\sqrt{1-H^2 r^2}} \left\{ 
\T_0 -\frac{q_0}{ \K} 
\left[ \frac{1}{r} + H\arctanh \left( Hr \right) \right] \right\} \,. 
\label{puredSsolution}
\ee
In this formal solution one must choose the integration constant $q_0$ to vanish to avoid a divergent temperature~(\ref{puredSsolution}) at $r=0$. This choice is physically well-motivated. In fact, in this case, the heat flux from the de Sitter horizon travels radially inward toward $r=0$, but it cannot accumulate there and must continue travelling outward to larger radii and cross the horizon, to never come back to the interior region. Ingoing and outgoing fluxes balance out, resulting in a zero net flux, and one must set $q_0=0$. As a check, in this case the solution~(\ref{puredSsolution}) gives back the Tolman\textendash{}Ehrenfest criterion~(\ref{standardTolman})
\be
\T (r) = \frac{ \T_0}{ \sqrt{1-H^2r^2}} \,.
\ee
This temperature diverges on the de Sitter horizon, as is well known.

\section{Temperature regularization at the horizons}
\label{sec:4}
\setcounter{equation}{0}

The divergence of the Tolman\textendash{}Ehrenfest temperature at the black hole horizon and its relation with the Hawking temperature are well known in standard horizon thermodynamics \cite{Birrell:1982ix, Wald:1999xu, Giddings:2015uzr}. Let us consider the Schwarzschild black hole: divergences are not new in thermodynamics and are related to the vacuum state of the quantum field used to compute the temperature (for example, the Boulware vacuum is singular on the black hole horizon). {The Hawking temperature is $\T_{\rm H} = \kappa / \left( 2\pi \right)$, where $\kappa$ is the surface gravity,\footnote{The surface gravity $\kappa$ is defined globally \cite{Wald:1984rg} instead of locally or quasilocally.} and it represents the temperature seen by a static distant observer ($r\to\infty$). To remain at fixed radius, a detector is accelerated with uniform acceleration $a$ and, in the inertial vacuum state, detects blackbody radiation at the Unruh temperature $\T_\mathrm{U}=a/\left( 2\pi \right)$.} In the Unruh vacuum the modes that, according to this observer, have propagated from the region near the black hole are populated (unlike the modes propagating from spatial infinity, which are not), at the temperature $\T=\kappa/\left( 2\pi \sqrt{-k^ak_a} \right)$, where $k^a$ is the local timelike Killing vector. As the Schwarzschild horizon is approached, in the limit $r\to r_\mathrm{S}=2m $, the surface gravity is 
obtained as \cite{Wald:1984rg,Birrell:1982ix} 
\be 
\kappa = \lim_{r\to 2m} \sqrt{-k^c k_c} 
\, a \,. 
\ee 
Therefore, 
\be 
\T =\frac{a}{2\pi} \to \frac{\kappa}{2\pi\sqrt{ -k^c k_c}} \,,
\ee 
and, since $ k^c k_c=g_{00} $ in coordinates adapted to the local time symmetry, 
\be 
\T \sqrt{-g_{00}} \simeq \frac{a}{2\pi} {\sqrt{-g_{00}}}
\simeq \frac{\kappa}{2\pi}\,
\ee 
for the Unruh observer. Therefore, the local temperature of the Hawking radiation obeys the Tolman\textendash{}Ehrenfest criterion (as remarked explicitly, e.g., in \cite{Birrell:1982ix, Wald:1999xu, Giddings:2015uzr}) and applying this criterion to the radiation fluid generated from the horizons, as we did in the previous section, is not far-fetched nor new.

There is an even more important aspect regarding the apparent divergence of the temperature $\T$. There is increasing consensus that the Hawking radiation does not arise at the horizon but in a quantum atmosphere surrounding it, which is macroscopic \cite{Giddings:2015uzr, Dey:2017yez, Dey:2019ugf, Ong:2020hti, Nambu:2021eqe, Kaczmarek:2023kpn}. That is, this atmosphere is not confined to a Planck-sized near the horizon, but it extends to radii $r\simeq \frac{3\sqrt{3}}{4} \, r_\mathrm{S} \simeq 1.3 r_\mathrm{S} $, as indicated by the fact that the Hawking flux indicates an emission area larger than the horizon area \cite{Casadio:2002dj,Giddings:2015uzr, Dey:2017yez, Dey:2019ugf, Ong:2020hti, Nambu:2021eqe, Kaczmarek:2023kpn}. Likewise, the wavelength of the typical Hawking quantum is not $r_\mathrm{S}$, but $\simeq 79 r_\mathrm{S}$ \cite{Birrell:1982ix, Giddings:2015uzr}. Presumably, similar considerations hold for the Gibbons\textendash{}Hawking radiation associated with the de Sitter horizon.

Another possibility for regularizing the divergence of the Tolman temperature consists of including the trace anomaly for the radiation \cite{Eune:2015xvx}. As a conclusion, the divergence of the temperature profile~(\ref{Tprofile}) can be eliminated in more nuanced descriptions of the relevant vacuum state and emission process, or is simply not occurring because of the macroscopic quantum atmosphere associated with horizons.

\section{Conclusions}
\label{sec:5}
\setcounter{equation}{0}

Thermal equilibrium (or the lack thereof) between the two horizons of the SdS spacetime exhibits some similarities with heat conduction in a rod of finite length whose ends are kept at constant, different, temperatures. 
The temperature gradient maintains a steady heat flow and there is thermal equilibrium in the sense that both the heat flux density $\vec{q}$ and the linear temperature profile $\T(x)$ are time-independent. In the SdS case, it is the blackbody radiation between the two horizons (a perfect fluid) that is in thermal equilibrium with a radial temperature profile $\T(r)$ and the heat flux density (which is purely spatial in Eckart's description). {This feature, which implies superluminal heat flow and is clearly unrealistic, is corrected in more refined causal thermodynamics formalisms \cite{Muller:1967zza, Stewart77, Israel:1979wp, IsraelStewart79b, HisckockLindblom99, Carter91, MullerRuggeri98}, but the Eckart model is still the simplest and widely adopted model of dissipative fluid in relativity \cite{Maartens:1996vi, Andersson:2006nr}.}

The SdS spacetime is locally static between the horizons, which remain at the Hawking and the Gibbons\textendash{}Hawking temperatures, respectively. A stationary heat flux with purely radial density $q^r (r) >0$ is established and the resultant temperature profile $ \T(r)$ is time-independent. Heat flows from the hotter black hole horizon to the cooler de Sitter horizon, from which it escapes forever the locally static region between these two causal boundaries. There is no breakdown of thermodynamic laws, as suggested in the literature, and the explanation proposed here of the two-temperature puzzle is both simple and economical in its assumptions. Our main point is that the horizons being at vastly different temperatures does not preclude thermal equilibrium in the SdS spacetime, as is often stated. Quite the opposite, the Tolman criterion of thermal equilibrium~(\ref{standardTolman}) in a static spacetime {\em requires} a temperature gradient: the temperature cannot be constant.

The main limitation of this work consists of the non-causal character of the dissipative fluid model adopted and of the heat flux. It remains an open question whether more refined, causal thermodynamic formalisms provide the same answer to the puzzle of the two temperatures of SdS and thermal equilibrium.

\begin{acknowledgments} 

M.~M. is grateful for the support of Istituto Nazionale di Fisica Nucleare (INFN) \textit{iniziativa specifica}  MOONLIGHT2. V.~F. is supported by the Natural Sciences \& Engineering Research Council of Canada (grant 2023-03234).

\end{acknowledgments}

\begin{appendices}

\section{Standard Tolman\textendash{}Ehrenfest criterion from 
Eckart's law}
\label{Appendix:A} 
\renewcommand{\theequation}{B.\arabic{equation}} 
\setcounter{equation}{0}

First, we need to prove the Buchdahl relation \cite{Buchdahl49} expressing the fact that in a static spacetime, in coordinates adapted to the time symmetry, the 4-acceleration of a particle is $a^a \equiv \dot{u}^a = \nabla^a \ln \left( \sqrt{-g_{00}} \, \right)$.

Let $k^a $ be the timelike Killing vector satisfying the Killing equation $\nabla_{(a} k_{b )} = 0 $. Its norm is conserved along the integral curve of $k^a$, $ k^a \nabla_a \left( g_{bc} k^b k^c \right) = 0 $, and $\nabla_a k^a = 0 $.

Let us consider the coordinate frame adapted to $k^a$, in which $k^c k_c=g_{00}$. An observer (e.g., a fluid element) following the Killing trajectory is at rest in these coordinates and has a 4-velocity 
\begin{equation}
 u^a = \frac{k^a}{ \sqrt{ -k^ck_c} } 
\end{equation}
and 4-acceleration
\begin{equation}
\dot{u}^a \equiv u^b \nabla_b u^a = -\frac{k^b \nabla_b k^a}{k^c k_c} \,.
\end{equation}
The Killing equation $\nabla_{(a}k_{b)} = 0 $ yields 
\begin{align} 
\dot{u}_a & = - \frac{ k^b\nabla_a k_b }{k^c k_c}= 
\frac{1}{2} \, \frac{\nabla_a \left( -k^b k_b \right)}{-k^c k_c} \nn\\
 & = \nabla_a \left( \ln \sqrt{ -k_b k^b} \right)= 
\nabla_a \left( \ln \sqrt{-g_{00} } \, \right) \,.
\end{align}

To complete the derivation of the Tolman\textendash{}Ehrenfest criterion from Eckart's law 
\be
 q^a = -\K \,h^{ab} \left( \nabla_b \T + \T \dot{u}_b 
\right) \,,
\ee
consider a viscous fluid with a stress-energy tensor of the form~(\ref{imperfect}). The spatial components of the fluid's 4-velocity and, consequently, the quantities describing dissipation vanish for a fluid at rest. In static coordinates the temperature $\T$ does not depend on time, $u^a \nabla_a \T = 0 $. Then, 
\be
h_{ab} \nabla^b \T = \left( g_{ab} + u_a u_b \right) \nabla^b \T 
= \nabla_a \T \,.
\ee 
The ``standard'' definition of thermal equilibrium corresponds to the absence of heat flow, $q^a=0$, in which case Eckart's law reads
\be
\nabla_a \T + \T \,\dot{u}_a = 0 \,.
\ee
The Buchdahl relation now gives
\be
\nabla_a \T + \T \,\nabla_a \left( \ln \sqrt{-g_{00}} \, 
\right) = 0 \,,
\ee
or
\be
\nabla_a \ln \left( \T \sqrt{-g_{00} } \, \right) = 0
\ee
yielding the ``standard'' Tolman\textendash{}Ehrenfest criterion
\be
\T \sqrt{-g_{00} } = \T_0 \,.
\ee

\section{Static spacetime and static fluid}
 \label{Appendix:B} 
\renewcommand{\theequation}{A.\arabic{equation}} 
\setcounter{equation}{0}

Static spacetimes are characterized by a timelike Killing vector and, in coordinates adapted to the time symmetry, the metric coefficients are time-independent, $\partial_t g_{\mu\nu}=0$ and $ g_{0i}=0$. In spherical symmetry, the metric is also diagonal.

A static fluid in this coordinate system has 4-velocity $u^a$ orthogonal to the hypersurfaces of constant time, with components 
\begin{equation}
 u^a = \frac{ {\delta^a}_0 }{\sqrt{-g_{00}}}\,.
\end{equation}
and the fluid quantities are time-independent,
\begin{equation}
 \partial_t \rho=u^c \nabla_c \rho \equiv \dot{\rho} = 0\,,
\end{equation}
\begin{equation} 
\Theta \equiv \nabla_c u^c = \frac{1}{\sqrt{-g}} \, 
\partial_c \left( \sqrt{-g}\, u^c \right) = \partial_t u^0 = 0\,.
\end{equation}
One also has $ {h^b}_a\dot{q}_b = 0 $, i.e., $ \dot{q}^a $ reduces to its time component in the static coordinates. 
In fact, $u^a = \left( u^0,0 ,0,0 \right) $ and $ q^a = \left( 0, q^i \right) $ with $\partial_t q^i = 0 $. Then,
\begin{align}
 \dot{q}^a\equiv u^{b}\nabla_b q^a = 
u^0 \left( \partial_t q^a + \Gamma^a_{0j} q^j \right) \,,
\end{align}
where the first term on the right-hand side vanishes because $q^i$ is time-independent and the second term vanishes because 
\begin{align}
 \Gamma^i_{0j}=\frac{1}{2} \, g^{ik}\left( 
\partial_t g_{jk} + \partial_j g_{0k} - \partial_k g_{j0} \right) = 0 \,.
\end{align}
Therefore,
\begin{align}
 \dot{q}^a = \left( \frac{1}{2} \, q^j u^0 g^{00} 
\partial_j g_{00}, 0, 0, 0 \right) = u^a q_b \dot{u}^b \,,\label{dotq}
\end{align}
where $\Gamma^0_{0j}=\frac{1}{2} \, g^{00}\partial_j g_{00} $.

Equation\z\eqref{dotq} can also be guessed by differentiating the relation $q^a u_{a}=0$: $u^{b}\nabla_{b}(q_{a}u^{a})=0 $ gives
\be 
q_a \dot{u}^a = -\dot{q}^a u_a \,.
\end{equation}

\end{appendices}


\begin{thebibliography}{99}

\bibitem{Hawking:1975vcx}
S.~W.~Hawking,
``Particle Creation by Black Holes,''
Commun. Math. Phys. \textbf{43}, 199-220 (1975)
[Erratum: Commun. Math. Phys. \textbf{46}, 206 (1976)]
doi:10.1007/BF02345020

\bibitem{Wald:1984rg}
R.~M.~Wald,
``General Relativity,''
Chicago Univ. Pr., 1984,
doi:10.7208/chicago/9780226870373.001.0001

\bibitem{Gibbons:1977mu}
G.~W.~Gibbons and S.~W.~Hawking,
``Cosmological Event Horizons, Thermodynamics, and Particle Creation,''
Phys. Rev. D \textbf{15}, 2738-2751 (1977)
doi:10.1103/PhysRevD.15.2738

\bibitem{Nariai}
 H. Nariai, ``On some static solutions of Einstein's gravitational 
field equations in a spherically symmetric case'', Sci. Rep. Tohoku Univ. 
{\bf 34}, 160 (1950)

\bibitem{Choudhury:2004ph}
T.~R.~Choudhury and T.~Padmanabhan,
``Concept of temperature in multi-horizon spacetimes: Analysis of 
Schwarzschild--de Sitter metric,''
Gen. Rel. Grav. \textbf{39}, 1789-1811 (2007)
doi:10.1007/s10714-007-0489-0
[arXiv:gr-qc/0404091 [gr-qc]].

\bibitem{Aros:2008ef}
R.~Aros,
``de Sitter Thermodynamics: A Glimpse into non equilibrium,''
Phys. Rev. D \textbf{77}, 104013 (2008)
doi:10.1103/PhysRevD.77.104013
[arXiv:0801.4591 [gr-qc]].

\bibitem{Saida:2009ss}
H.~Saida,
``To what extent is the entropy-area law universal?: Multi-horizon and 
multi-temperature spacetime may break the entropy-area law,''
Prog. Theor. Phys. \textbf{122}, 1515-1552 (2010)
doi:10.1143/PTP.122.1515
[arXiv:0910.2510 [gr-qc]].

\bibitem{Kim:2014zta}
S.~W.~Kim,
``The Hawking temperature of a dynamical black hole in de Sitter 
spacetime,''
Grav. Cosmol. \textbf{20}, no.4, 247-251 (2014)
doi:10.1134/S0202289314040094

\bibitem{Pappas:2017kam}
T.~Pappas and P.~Kanti,
``Schwarzschild\textendash{}de Sitter spacetime: The role of temperature 
in the emission of Hawking radiation,''
Phys. Lett. B \textbf{775}, 140-146 (2017)
doi:10.1016/j.physletb.2017.10.058
[arXiv:1707.04900 [hep-th]].

\bibitem{Robson:2019yzx}
C.~W.~Robson, L.~D.~Villari and F.~Biancalana,
``Global Hawking Temperature of Schwarzschild--de Sitter Spacetime: a 
Topological Approach,''
[arXiv:1902.02547 [gr-qc]].

\bibitem{Nakarachinda:2021jxd}
R.~Nakarachinda, E.~Hirunsirisawat, L.~Tannukij and P.~Wongjun,
``Effective thermodynamical system of Schwarzschild\textendash{}de Sitter 
black holes from R\'enyi statistics,''
Phys. Rev. D \textbf{104}, no.6, 064003 (2021)
doi:10.1103/PhysRevD.104.064003
[arXiv:2106.02838 [gr-qc]].

\bibitem{Volovik:2023stf}
G.~E.~Volovik,
``On the Global Temperature of the Schwarzschild\textendash{}de Sitter 
Spacetime,''
JETP Lett. \textbf{118}, no.1, 8-13 (2023)
doi:10.1134/S0021364023601173
[arXiv:2304.09847 [gr-qc]].

\bibitem{Tolman28} R. C. Tolman, ``On the extension of thermodynamics to 
general relativity'', Proceedings of the National Academy of Sciences of 
the United States of America, 14 \# 3 (1928), pp. 268-272. 
http://www.jstor.org/stable/85423

\bibitem{Tolman:1930zza} R.~C.~Tolman, ``On the Weight of Heat and Thermal 
Equilibrium in General Relativity,'' Phys. Rev. \textbf{35}, 904-924 
(1930) doi:10.1103/PhysRev.35.904

\bibitem{Tolman:1930ona} R.~Tolman and P.~Ehrenfest, ``Temperature 
Equilibrium in a Static Gravitational Field,'' Phys. Rev. \textbf{36}, 
no.12, 1791-1798 (1930) doi:10.1103/PhysRev.36.1791

\bibitem{MTW} C. Misner, K.S. Thorne, and J.A. Wheeler, {\em Gravitation} 
(Freeman, San Francisco, 1973), exercise~22.7, p.~567

\bibitem{Santiago:2018kds} J.~Santiago and M.~Visser, ``Tolman 
temperature 
gradients in a gravitational field,'' Eur. J. Phys. \textbf{40}, no.2, 
025604 (2019) doi:10.1088/1361-6404/aaff1c [arXiv:1803.04106 [gr-qc]].

\bibitem{Santiago:2019aem} J.~Santiago, ``On the Connections between 
Thermodynamics and General Relativity,'' PhD thesis, Victoria University 
of Wellington, New Zealand, 2019 [arXiv:1912.04470 [gr-qc]].

\bibitem{Santiago:2018jeu} J.~Santiago and M.~Visser, 
``Gravity\textquoteright{}s universality: The physics underlying Tolman 
temperature gradients,'' Int. J. Mod. Phys. D \textbf{27}, no.14, 1846001 
(2018) doi:10.1142/S021827181846001X [arXiv:1805.05583 [gr-qc]].

\bibitem{Eckart40_I} C. Eckart, ``The Thermodynamics of Irreversible 
Processes. I. The Simple Fluid,'' Phys. Rev. \textbf{58}, 267 (1940) 
doi:10.1103/PhysRev.58.267.

\bibitem{Eckart:1940te} C.~Eckart, ``The thermodynamics of irreversible 
processes. 3.~Relativistic theory of the simple fluid,'' Phys. Rev. 
\textbf{58}, 919-924 (1940), doi:10.1103/PhysRev.58.919

\bibitem{Klein49} O. Klein, ``On the thermodynamical equilibrium of fluids 
in gravitational fields'', Rev. Mod. Phys. \textbf{21}, 531 (1949).

\bibitem{Luttinger:1964zz} J.~M.~Luttinger, ``Theory of Thermal Transport 
Coefficients,'' Phys. Rev. \textbf{135}, A1505-A1514 (1964) 
doi:10.1103/PhysRev.135.A1505

\bibitem{Laskos-Patkos:2022lgy} P.~Laskos-Patkos, P.~S.~Koliogiannis, 
A.~Kanakis-Pegios and C.~C.~Moustakidis, ``Thermodynamics of Hot Neutron 
Stars and Universal Relations,'' Universe \textbf{8}, no.8, 395 (2022) 
doi:10.3390/universe8080395 [arXiv:2207.03347 [astro-ph.HE]].

\bibitem{Li:2022url} J.~Li, T.~Guo, J.~Zhao and L.~He, ``Do we need dense 
matter equation of state in curved spacetime for neutron stars?,'' Phys. 
Rev. D \textbf{106}, no.8, 083021 (2022) doi:10.1103/PhysRevD.106.083021 
[arXiv:2206.02106 [gr-qc]].

\bibitem{Kim:2021kou} H.~C.~Kim and Y.~Lee, ``Local temperature in general 
relativity,'' Phys. Rev. D \textbf{105}, no.8, L081501 (2022) 
doi:10.1103/PhysRevD.105.L081501 [arXiv:2110.00209 [gr-qc]].

\bibitem{Kim:2022qlc} H.~C.~Kim and Y.~Lee, ``Heat conduction in general 
relativity,'' Class. Quant. Grav. \textbf{39}, no.24, 245011 (2022) 
doi:10.1088/1361-6382/aca1a1 [arXiv:2206.09555 [gr-qc]].

\bibitem{Lima:2019brf} J.~A.~S.~Lima, A.~Del Popolo and A.~R.~Plastino, 
``Thermodynamic Equilibrium in General Relativity,'' Phys. Rev. D 
\textbf{100}, no.10, 104042 (2019) doi:10.1103/PhysRevD.100.104042 
[arXiv:1911.09060 [gr-qc]].

\bibitem{Lima:2021ccv} J.~A.~S.~Lima and J.~Santos, 
``Tolman-Ehrenfest-Klein law in non-Riemannian geometries,'' Phys. Rev. D 
\textbf{104}, no.12, 124089 (2021) doi:10.1103/PhysRevD.104.124089 
[arXiv:2112.12282 [gr-qc]].

\bibitem{Faraoni:2023gqg} V.~Faraoni and R.~Vanderwee, 
``Tolman-Ehrenfest\textquoteright{}s criterion of thermal equilibrium 
extended to conformally static spacetimes,'' Phys. Rev. D \textbf{107}, 
no.6, 064072 (2023) doi:10.1103/PhysRevD.107.064072 [arXiv:2301.09021 
[gr-qc]].

\bibitem{Karolinski:2024ukr}
N.~Karolinski and V.~Faraoni,
``The Tolman-Ehrenfest criterion of thermal equilibrium in scalar-tensor 
gravity,''
[arXiv:2406.09251 [gr-qc]].

\bibitem{Ellis71} G.~F.~R.~Ellis, ``Relativistic cosmology,'' Proc. Int. 
Sch. Phys. Fermi \textbf{47} (1971), 104-182 doi:10.1007/s10714-009-0760-7

\bibitem{Buchdahl49} H. A. Buchdahl, ``Temperature Equilibrium in a 
Stationary Gravitational Field'', Phys. Rev. \textbf{76}, 427-428 (1949).


\bibitem{Casadio:2002dj}
R.~Casadio and L.~Mersini-Houghton,
``Short distance signatures in cosmology: Why not in black holes?,''
Int. J. Mod. Phys. A \textbf{19}, 1395-1412 (2004)
doi:10.1142/S0217751X04016453
[arXiv:hep-th/0208050 [hep-th]].

\bibitem{Giddings:2015uzr}
S.~B.~Giddings,
``Hawking radiation, the Stefan\textendash{}Boltzmann law, and 
unitarization,''
Phys. Lett. B \textbf{754}, 39-42 (2016)
doi:10.1016/j.physletb.2015.12.076
[arXiv:1511.08221 [hep-th]].

\bibitem{Dey:2017yez}
R.~Dey, S.~Liberati and D.~Pranzetti,
``The black hole quantum atmosphere,''
Phys. Lett. B \textbf{774}, 308-316 (2017)
doi:10.1016/j.physletb.2017.09.076
[arXiv:1701.06161 [gr-qc]].

\bibitem{Dey:2019ugf}
R.~Dey, S.~Liberati, Z.~Mirzaiyan and D.~Pranzetti,
``Black hole quantum atmosphere for freely falling observers,''
Phys. Lett. B \textbf{797}, 134828 (2019)
doi:10.1016/j.physletb.2019.134828
[arXiv:1906.02958 [gr-qc]].

\bibitem{Ong:2020hti}
Y.~C.~Ong and M.~R.~R.~Good,
``Quantum atmosphere of Reissner-Nordstr\"om black holes,''
Phys. Rev. Res. \textbf{2}, no.3, 033322 (2020)
doi:10.1103/PhysRevResearch.2.033322
[arXiv:2003.10429 [gr-qc]].

\bibitem{Nambu:2021eqe}
Y.~Nambu and S.~Noda,
``Interferometry of black holes with Hawking radiation,''
Phys. Rev. D \textbf{105}, no.4, 045022 (2022)
doi:10.1103/PhysRevD.105.045022
[arXiv:2109.07044 [gr-qc]].

\bibitem{Kaczmarek:2023kpn}
A.~Z.~Kaczmarek and D.~Szcz\k{e}\'sniak,
``Signatures of the black hole quantum atmosphere in nonlocal 
correlations,''
Phys. Lett. B \textbf{848}, 138364 (2024)
doi:10.1016/j.physletb.2023.138364
[arXiv:2306.09941 [gr-qc]].

\bibitem{Eune:2015xvx}
M.~Eune, Y.~Gim and W.~Kim,
``Effective Tolman temperature induced by trace anomaly,''
Eur. Phys. J. C \textbf{77}, no.4, 244 (2017)
doi:10.1140/epjc/s10052-017-4812-y
[arXiv:1511.09135 [gr-qc]].


\bibitem{Birrell:1982ix}
N.~D.~Birrell and P.~C.~W.~Davies,
``Quantum Fields in Curved Space,''
Cambridge Univ. Press, 1984, pp.~109-117 
ISBN 978-0-521-27858-4, 978-0-521-27858-4
doi:10.1017/CBO9780511622632

\bibitem{Wald:1999xu} R.~M.~Wald, ``Gravitation, thermodynamics, and 
quantum theory,'' Class. Quant. Grav. \textbf{16}, A177-A190 (1999) 
doi:10.1088/0264-9381/16/12A/309 [arXiv:gr-qc/9901033 [gr-qc]].

\bibitem{Muller:1967zza}
I.~Muller,
``Zum Paradoxon der Warmeleitungstheorie,''
Z. Phys. \textbf{198}, 329-344 (1967)
doi:10.1007/BF01326412

\bibitem{Stewart77} J.~M.~Stewart, ``On transient relativistic 
thermodynamics and kinetic theory,'' Proc. R. Soc. London, Ser. A {\bf 
357} (1977) 59

\bibitem{Israel:1979wp}
W.~Israel and J.~M.~Stewart,
``Transient relativistic thermodynamics and kinetic theory,''
Annals Phys. \textbf{118}, 341-372 (1979)
doi:10.1016/0003-4916(79)90130-1

\bibitem{IsraelStewart79b} W.~Israel and J.~M.~Stewart, ``On transient 
relativistic thermodynamics and kinetic theory.~II,'' Proc. R. Soc. 
London, Ser. A {\bf 365} (1979), 43-52, doi:10.1098/rspa.1979.0005

\bibitem{HisckockLindblom99} W.~A.~Hiscock and L.~Lindblom, ``On transient 
relativistic thermodynamics and kinetic theory.~II,'' Phys. Lett. A {\bf 
131} (1988) 509

\bibitem{Carter91} B.~Carter, ``Convective variational approach to 
relativistic thermodynamics of dissipative fluids,'' Proc. R. Soc. London, 
Ser. A {\bf 433} (1991) 45, doi:10.1098/rspa.1991.0034

\bibitem{MullerRuggeri98} I. M\"uller and T. Ruggeri, {\em Rational 
Extended Thermodynamics}, 2nd edition (Springer, New York, 1998)

\bibitem{Maartens:1996vi} R.~Maartens, ``Causal thermodynamics in 
relativity,'' [arXiv:astro-ph/9609119 [astro-ph]].

\bibitem{Andersson:2006nr} N.~Andersson and G.~L.~Comer, ``Relativistic 
fluid dynamics: Physics for many different scales,'' Living Rev. Rel. 
\textbf{10}, 1 (2007), doi:10.12942/lrr-2007-1 [arXiv:gr-qc/0605010 
[gr-qc]].

\end{thebibliography}
\end{document}